\documentstyle[twocolumn,prb,aps,floats,epsf,lscape]{revtex}\begin{document}
\draft
\twocolumn[\hsize\textwidth\columnwidth\hsize\csname@twocolumnfalse\endcsname
\vspace{-1cm}
\begin{flushright}
Submitted to PHYSICAL REVIEW B15 on 16 April 1998
\end{flushright}
\title
{\bf Quantum point contact on graphite surface}
\author{\c{C}.~K{\i}l{\i}\c{c}, H.~Mehrez, and S.~Ciraci}
\address{
Department of Physics, Bilkent University, Bilkent 06533,
Ankara, Turkey.}
\maketitle
\begin{abstract}
The conductance through a quantum point contact created by a sharp and hard metal tip on the graphite surface has features which to our knowledge have not been encountered so far in metal contacts or in nanowires.
In this paper we first investigate these features which emerge from the strongly directional bonding and electronic structure of graphite, and provide a theoretical understanding for the electronic conduction through quantum point contacts.
Our study involves the molecular-dynamics simulations to reveal the variation of interlayer distances and atomic structure at the proximity of the contact that evolves by the tip pressing toward the surface.
The effects of the elastic deformation on the electronic structure, state density at the Fermi level, and crystal potential are analyzed by performing self-consistent-field pseudopotential calculations within the local-density approximation.
It is found that the metallicity of graphite increases under the uniaxial compressive strain perpendicular to the basal plane.
The quantum point contact is modeled by a constriction with a realistic potential.
The conductance is calculated by representing the current transporting states in Laue representation, and the variation of conductance with the evolution of contact is explained by taking the characteristic features of graphite into account.
It is shown that the sequential puncturing of the layers characterizes the conductance.
\pacs{73.40.Cg, 73.23.Ad, 62.20.Dc}
\end{abstract}
\vskip1pc]
\section{Introduction}\label{sone}
Graphite was a prototype sample in early experiments\cite{binnig85} aiming at atomic resolution in scanning tunneling microscopy (STM), since atomically flat and perfect surfaces  over thousands of angstroms can be achieved in air.
Despite this convenient situation  some STM data obtained from the graphite surface were rather puzzling, and have been the focus of interest.
For example, that only three alternating atoms, out of six atoms of hexagons in a honeycomb structure can be imaged demonstrated the crucial role of the sample electronic structure in STM.\cite{selloni84,batra87,tomanek87}
The giant corrugation\cite{soler86} recorded at a small tip-sample distance came out rather unexpectedly, and has pointed out the importance of tip-sample interaction effects\cite{ciraci87,ciraci89a,ciraci90} in STM.
Atomically resolved images of graphite with corrugation up to $24$~\AA~ were obtained in air while the tip progressively displaced over $100$~\AA~.
It was argued that the contamination under the tip distributes the perpendicular tip force over a large area, and that the elastic deformation of the tip amplifies the corrugation.\cite{soler86,pethica86}
Moreover, the conductance of the atomic size contact and its variation with the perpendicular displacement $s$ of the tip have been rather confusing,\cite{oral} and yet cannot be reconciled with what has been learned from metallic contacts.\cite{gimzewski87,lang87,ferrer88,ciraci89b}

Experiments\cite{gimzewski87} on a metal sample (such as Ag) showed that the conductance versus the perpendicular displacement of the tip, $G(s)$, is initially very small, but increases exponentially because of tunneling through the vacuum gap between the tip and sample surface.
Usually, it first saturates at $G <2e^2 n/h$ upon the onset of contact,\cite{ciraci87} then increases discontinuously with the growth of the contact.
Earlier, based on the first calculation of conductance of a three-dimensional, constriction the abrupt changes of $G(s)$ were attributed to the discontinuous change of the cross section of contact $A_c(s)$.\cite{ciraci89b}
Recent experiments\cite{agrait95,stalder96} achieving simultaneous measurements of the perpendicular tip force $F_z(s)$ and conductivity $G(s)$, as well as extensive molecular dynamics simulations,\cite{sutton90,todorov93,buldum98,mehrez97} have confirmed this argument.
The question of whether the conductance in an atomic size contact or in a nanowire (produced by retracting the tip from an extensive indentation) is quantized is a subject of current interest.\cite{refree}

The contact is usually set by a single atom at the apex, but it grows by additional tip atoms engaging in contact with the metal sample.\cite{todorov93,buldum98}
Initially, the contact radius $R_c$ is small, and $\sim2-5$  \AA; it is in the range of the Fermi wavelength $\lambda_F$ of metals.
For alkali metals, even metals having a half-filled $s$ band (Au, Cu, etc.), the concepts and scales adopted from a free-electron gas (FEG), such as spherical Fermi surface, $\lambda_F$, $E_F$, etc., can be used to characterize the contact and its conductance.
Under these circumstances, once $R_c \sim \lambda_F$, the level spacing of electrons (transversally confined to the contact) can be in the $eV$ range; such a quantization of electronic motion in the constriction is reflected by ballistic electron transport through the contact, even at room temperature.
In developing the theory of conductance through a metal contact, FEG systems with metallic densities were usually postulated. 
It was also implicitly assumed that the tip and sample have the same metal.
Under these assumptions the conductance can be calculated by representing the contact by a finite (but short)  constriction in which the potential is uniform.
For such a model, even Sharvin's formula\cite{sharvin65} $G_S=(2e^2/h)(\pi R_c/\lambda_F)^2$ is able to describe the overall behavior of conduction, except for features arising from detailed atomic structure.
That such a description of a metal contact cannot always be valid, and hence the nature of atoms at the contact may be crucial, becomes evident in the ferromagnetic nanowires.\cite{hansen97,mehrez98a}
In fact, the self-consistency of the potential, the mismatch of Fermi surfaces between electrodes (tip and sample) that are made of markedly different metals, and conservation of momentum have not been treated thoroughly yet.

\begin{figure}
\vspace{-3.5cm}
\epsfxsize=5.25 truein
\centerline{\hspace{0.75cm}\epsfbox{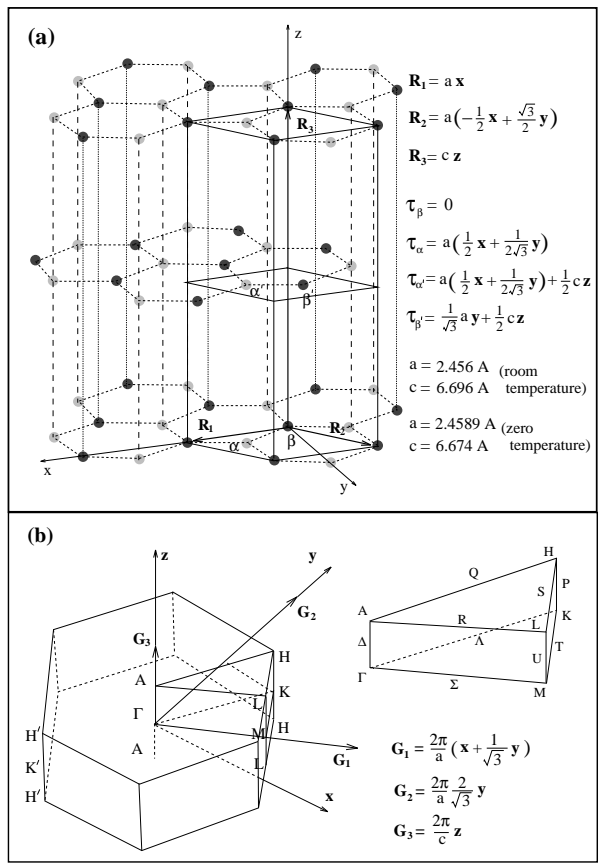}}
\vspace{-3cm}
\caption{(a) The atomic arrangement and the unit cell of the Bernal graphite. 
The equilibrium lattice parameters at zero temperature and at room temperature are taken from Refs.~28~and~29, respectively.
(b) The first Brillouin zone and its irreducible wedge with symmetry points and directions. }
\vspace{-.25cm}
\end{figure}

The situation in graphite is even more complex and rather different from metal contacts.
From the electronic structure point of view, graphite is a semimetal with complex and narrow Fermi surface along the $(HKH)$ edges of the hexagonal Brillouin zone (BZ).
Energy bands contributing to the Fermi surface are rather flat, and the free-carrier density is rather low.
From the atomic structure point of view, graphite exhibits a strongly directional bonding structure that leads to atomic planes (basal planes) with wide interlayer spacings.
Owing to the weak interlayer coupling the crystal potential at the interlayer region is only $\sim2$eV below the Fermi level.
As a result the transport and mechanical properties (elastic stiffness tensor, phonon spectrum) exhibit a strongly directional behavior.
The evolution of atomic structure during the growth of contact, and the related variation of conductance, involve several features different from a metal contact.
So far the tunneling between a metal tip and a graphite surface in STM has been treated only qualitatively, yet no quantitative study has been provided.

This paper presents a thorough analysis\cite{kilic97} of the formation and growth of the contact on a (0001) surface (or basal plane) of graphite, and provides a theoretical understanding of electron transport through this contact by taking into account characteristic features which do not exist in usual metal contacts.
To this end, the deformation of layers and the evolution of an atomic structure with the tip pressing against the surface is calculated by using classical molecular dynamics with an empirical potential\cite{nordlund96} derived from Tersoff's many-body potential.\cite{tersoff86}
In this work, this potential is further elaborated upon to work better under high uniaxial strain.
The effect of the tip induced deformation is revealed by the self-consistent-field (SCF) pseudopotential calculation of the total energy, electronic band structure, and total density of states as functions of the lattice parameter $c$.
The results of atomic simulations and {\it ab initio} calculation of the electronic structure and crystal potential are combined in a model constriction, whereby the variation of conductance is analyzed by calculating the current within the linear response theory.

\section{Atomistic Simulations}\label{stwo}
Each carbon atom with its three $sp^2$ hybrid orbitals is attached to the three nearest-neighbor C atoms in the same plane. 
This way C atoms are arranged in a honeycomb structure in the $xy$ plane and form the individual $(0001)$ basal plane, or graphene.   
The graphite crystal forms by stacking of graphenes along the $z$ direction (or $[0001]$ direction) with an equilibrium separation $d=3.348$~\AA~at room temperature.
In the normal stacking sequence (of Bernal graphite) one plane is shifted relative to the adjacent plane, so that three alternating ($\alpha$) atoms of a hexagon directly face three atoms ($\alpha^\prime$) of the adjacent graphenes.
Accordingly, remaining three ($\beta$) atoms face the centers of the hexagons (i.e. the H sites) in the adjacent graphenes.
This causes ABABAB stacking.
The atomic arrangements, lattice parameters, and the first BZ with symmetry points are described in Fig.~1.
In rhombohedral graphite, the layer stacking is ABCABC.
The AAA sequence of graphenes occurs in the first stage intercalates.
Owing to the strong bonding combination, of $sp^2$ orbitals between two nearest-neighbor C atoms, the C-C distance of a hexagon is $\sim1.418$~\AA; it is even shorter than the C-C distance in diamond structure.
The cohesive energy is $\sim7$~eV/atom\cite{brandt88} and the strength is rather high within the graphene.
In this respect the graphene is an essential unit of graphite. 
On the other hand, the interplanar interaction and resulting binding energy is weak and occurs  through the small overlap $<p_z|H|p_{z^\prime}>$ between the atoms of the adjacent layers and partly through the long range van der Waals interaction.
Pseudopotential calculations\cite{tomanek87,furthfuller94} predicts total-energy differences smaller than $5$ meV/atom among three types of layer sequences (ABAB..., ABCABC..., AAAA...)
The interlayer binding interaction (the experimental exfoliation energy) is only $22.8$ meV/atom.\cite{girifalco56}
The present and previous\cite{schabel92} pseudopotential calculations yield very close energies for interlayer binding.
In Sec.~\ref{sthree}, we show how these interactions are reflected to the electronic properties.

The empirical potential used in our molecular dynamics simulations starts from the form extended to multilayer graphite inRef.~26 by combining three distinct potentials:
\begin{eqnarray}
V(R_{ij})=[V_T(R_{ij})&+&V_G(R_{ij})]F(R_{ij})\nonumber\\
&+&V_R(R_{ij})[1-F(R_{ij})].\label{eone}
\end{eqnarray}
Here $R_{ij}$ is the distance between the $i$th and $j$th atoms.
The total energy of atomic system for a given configuration is then expressed by $E_T=\frac{1}{2}\sum_{i\ne j} V(R_{ij})$.
The main contribution to $V(R_{ij})$ is Tersoff's potential\cite{tersoff86} $V_T(R_{ij})$, that yields a good description of the bonding in diamond and graphene.
Owing to the relatively shorter range, the weak interaction between layers are not included in $V_T$.
The potential $V_G$ is introduced to include interlayer interaction, and a strong repulsive potential $V_R$ prevents the layers from collapsing.
The Fermi function $F(R_{ij})$ provides a smooth transition from the many-body combination $V_T(R_{ij})+V_G(R_{ij})$ to the repulsive potential $V_R(R_{ij})$. 
The above description is similar to that of the pair potential, however, there is an important difference due to the implicit many-body interactions in $V_T$ and $V_G$.
Tersoff went beyond the conventional two- and three-body potentials in transferability and accuracy by introducing a new scheme in view of the quantum-mechanical arguments brought about by the universal binding energy (Rydberg) curve of Rose and co-workers.\cite{rose81}
Furthermore, following Abell,\cite{abell85} Tersoff incorporated the bond order as depending on the local atomic environment in the empirical potential.
To avoid instabilities under excessive uniaxial strain, we made the following improvement for the potential  given in Eq.~(\ref{eone}).
Owing to the diamond-to-graphite transition data in its construction, the potential normally ``chooses'' the tetrahedral local environment rather than the graphitic coordination under the influence of high pressure.
However, in the case of uniaxial compression, there appears to be no interlayer interaction  if the interlayer spacing is between $2.46$ and $2.87$~\AA, because $V_T(R_{ij})$ and $V_G(R_{ij})$ are identically zero when $R_{ij} >2.46$~\AA~ and $R_{ij}< 2.87$~\AA, respectively; and the range of $V_R$ is even shorter than that of $V_T$.
For this reason, the potential results in a flat region between $c=4.92$ and $5.74$~\AA, as seen in the inset of Fig.~2(a), and the slope gained by the Tersoff potential remains too small in comparison with the experimental curve.\cite{divincenzo83}
We made a linear interpolation of $V_T$ and $V_G$ from $R_{ij}=1.8$~\AA~ to $R_{ij}=2.87$~\AA~ by keeping the parameters\cite{nordlund96,tersoff86} unchanged. 
This way the range of $V_G$ is extended toward that of $V_T$, and hence a weak barrier is added to $V(R_{ij})$ within the interpolation interval.
By this improvement the interlayer interaction curve becomes closer to that obtained by fitting to the experiment as shown in Fig.~2(a).
The transferability of the potential is not destroyed, and is even slightly improved for high-coordinated structures when compared to {\it ab initio} data,\cite{yin83} as seen in Fig.~2(b).

The formation of an atomic size contact on the graphite (0001) surface is simulated by using the classical molecular-dynamics method with the empirical potential explained above. 
In order to study the effect of the size on the results of simulations the tip-sample system is represented by three models of different size.
The first one comprises six layers, and the total number of sample atoms is 2016.
The second (third) model is comprises eight layers, corresponding to 3584 (4608) total number of sample atoms.
The positions of the atoms in the last two layers are kept fixed.
\begin{figure}
\vspace{-3cm}
\epsfxsize=5 truein
\centerline{\hspace{0.75cm}\epsfbox{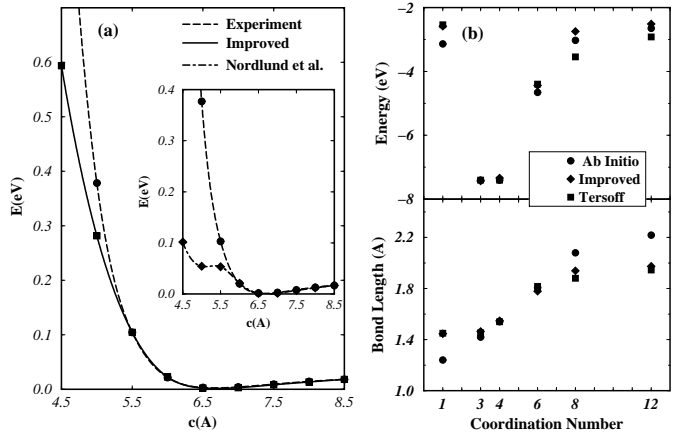}}
\vspace{-8.5cm}
\caption{(a) Comparison of the experimental total-energy variation with the improved potential.
In the inset a similar comparison is made with the potential given in Ref.~26.
(b) The comparison of the transferability of the interatomic carbon potential with {\it ab initio} data. 
The values on the horizontal axis represents dimer (1), graphite (3), diamond (4), simple cubic (6), body-centered cubic (8), and face-centered cubic (12) structures.}
\vspace{-.25cm}
\end{figure}
The hard, sharp metal tip is represented by a robust diamond tip; it comprises 13 (111) diamond planes, and contains 167 carbon atoms.
The apex of the tip has a single atom, and the following layers contain 3, 3, 6, 6, 10, 10, 15, 15, 21, 21, 28, and 28 atoms. 
The $z$ axis is taken perpendicular to the (0001) surface and the periodic boundary conditions are imposed in the $xy$ plane.
The temperature is rescaled to 2 K at every two steps to avoid possible divergences in the kinetic energy of moving atoms.
The time step $\tau$ is taken to be $10^{-16}$ s, in accordance with high-order Gear algorithm.\cite{allen87}
Initially the sample is equilibrated in $\sim500$ relaxation steps before the tip starts to be pushed down from the height $h=2.5$~\AA.
As $h$ decreases, the number of initial relaxation steps increases; the equilibration is terminated when the fluctuations in the total energy is settled down.
After the initial relaxation stage, the tip is pushed at a rate of $1\times 10^{-4}$~\AA~ per time step for $500$ steps, and then the system is relaxed during the following $500$ steps.
Even if the velocity of the tip is faster than the experiment, it is small enough to allow the system to be equilibrated between successive instabilities if any occur.
That the average speed of the tip, $v=50$ m/s is appropriate, and the results of simulation are converged are tested by performing calculations with different pushing speeds and by analyzing the variation of temperature and potential energy of the whole system.

\begin{figure}
\vspace{-3cm}
\epsfxsize=4.5 truein
\centerline{\hspace{0.5cm}\epsfbox{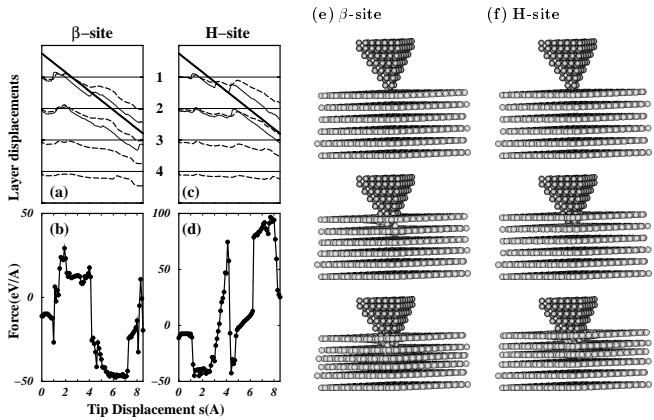}}
\vspace{-7cm}
\caption{(a) Variation of the layer heights with the tip displacement $s$.
(b)~corresponding force variation. The tip is positioned above $\beta$ site.
(c) and (d) are the same for the H site.
The dashed curves indicate the layer heights averaged over the heights of the atoms in the same layer of graphite.
The solid curves correspond to the averaging at close proximity of the tip.
The thick line is the position of the apex of the tip.
(e)~Snapshots of the evolution of contact at the $\beta$ site.
(f)~The same for the H site. }
\vspace{-.25cm}
\end{figure}

To examine the formation and evolution of the contact the tip is located at various special positions ($\alpha$, $\beta$, and H site) on the graphite surface, and it is pushed toward the surface (along the $z$ axis) by keeping its $(x,y)$ position fixed.
The effect of the tip displacement and resulting perpendicular tip-sample force are displayed in Fig.~3 for different lateral positions of the tip.
In the top panels are illustrated the averaged change of the layer heights with the tip displacement.
For the surface and subsurface layers the average change of heights only at a close proximity to the tip are also shown.
The bottom panels show the force variation $F_z(s)$. 
At the beginning, $h=2.5$ \AA; the tip-sample force is attractive in both cases.
However, when $s=1$ \AA, for the $\beta$ site the force enters into the repulsive range and increases with increasing $s$.
The repulsive range, in which the layers are compressed and hence the interlayer distance are decreased locally, last until $s\simeq 4$ \AA~ [see Fig.~3(b)].
At that point $F_z(s)$ drops suddenly but becomes attractive due to the puncture of the first graphite layer under high local pressure.\cite{abraham89}
The attractive force, that originates mainly from the attraction between the atoms at the side of the tip and those on the surface, coexists in the course of the push, but becomes dominant only after the release of the repulsive force at the apex following the puncture.
The strength of the attractive interaction of the metal tip cannot be as strong as that of the diamond tip.
Note that the decrease of the interlayer distance could be much smaller than $s$ and so the puncture would have occurred for relatively larger $s$ if the graphite slab under study had included more layers.
For $4$ \AA~$<s\le 7$ \AA, $F_z(s)$ stays in the attractive range but changes to the repulsive range for $s>8$ \AA, where the strong repulsive forces between the apex and second layer start to dominate the existing attractive forces.
This behavior continues periodically.
We note that the force variation for the H site in Fig.~3(d) takes place in reverse order relative to that of the $\beta$ site. 
This is due to the fact that in the Bernal graphite the $\beta$ and H sites of graphite planes occur alternately along the $z$ axis.
\begin{figure}
\vspace{-2.75cm}
\epsfxsize=2.5 truein
\centerline{\hspace{0.3cm}\epsfbox{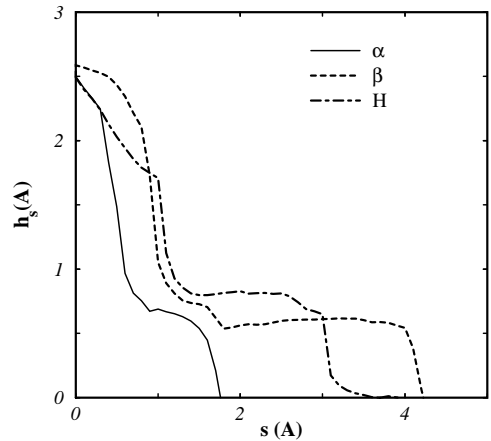}}
\vspace{0cm}
\caption{Variation of the tip-surface separation (calculated at the close proximity of the tip) with the tip displacement}
\vspace{-.25cm}
\end{figure}
Therefore, when the tip starts at the $\beta$ site of the first layer, it faces the H site of the second layer following the puncture of the first one.
The important results obtained from these atomistic simulations are summarized:
With increasing $s$, the tip first presses the layers.
This causes the interlayer distance to decrease locally.
Once $F_z(s)$ reaches a threshold value, the first layer is punctured, whereby the repulsive force is released and the graphite layers under compression are relaxed to maintain their interlayer distance, temporarily.
However, since $s$ continues to increase the compression of graphite starts again and the same events repeat quasiperiodically, i.e. compression of layers, puncture of the layer under the tip, and relaxation.
The evolution of the contact is described by the snapshots of the atomic structure obtained from simulations in Figs.~3(e)~and~3(f).
Note that the above behavior is different from the contact formed on the metal surfaces, where the quasiperiodic cycles, i.e. compression-relaxation followed by the puncture of layer is absent; the contact grows by the implementation of atoms to the contact area.\cite{sutton90,todorov93,buldum98}
While the puncture of the atomic plane occurs through the breaking of the bonds at $\beta$ and $H$ sites, the puncture initiate the formation of flakes if the contact is set at the $\alpha$ site.

Another interesting feature that is absent in metal contacts is the jump to contact of the graphite surface towards the tip.
As the tip approaches the sample the surface atoms at close proximity of the tip first move toward the tip and then maintain the separation $h_s$ approximately unaltered for a significant displacement of the tip even if the interlayer distance decreases. (see Fig.~4).
This situation lasts until the plastic deformation sets in.
The attraction of the atoms under the tip resulting in the jump to contact depends on the position of the tip ($\alpha$, $\beta$, and H sites) and its orientation relative to the honeycomb structure.

\section{Electronic Structure Calculations}\label{sthree}
Effects of the tip-induced deformation on the electronic energy structure and electronic potential of the contact are investigated by using SCF pseudopotential calculations in momentum space within the local-density approximation and by using the Ceperley-Alder exchange-correlation potential.\cite{ceperley80}
The ionic potential of carbon is represented by nonlocal, norm-conserving pseudopotential.\cite{bachelet82}
The tip-induced deformation is represented by a uniaxial strain and total energy, band structure $E({\bf k})$, local density of states $\rho(E,{\bf r})$, total density of states $D(E)$, and crystal potential $V_c({\bf r})$ are calculated for different interlayer distances. 
The kinetic energy cutoff $(\hbar^2/2m)|{\bf k}+{\bf G}|^2$ is taken to be $37$ Ry that corresponds to approximately $900$ plane waves for equilibrium structure.
The irreducible wedge of the BZ is sampled by uniformly distributed $48$ $\bf k$ points.
The convergences relative to the plane wave basis set and the $\bf k$-point sampling in BZ are tested by repeating the calculations with $33~Ry<\frac{\hbar^2}{2m}|{\bf k}+{\bf G}|^2<45~Ry$ and with uniform mesh points $216,~360,~432$.
The change in the total energy is found to be smaller than $0.5 \%$ in each case.
Taking into account the accuracy achieved by standard LDA calculations\cite{kohn83}, the kinetic-energy cutoff and $\bf k$-point sampling used in the present calculations are found to be appropriate to reveal the effect of deformation on the electronic transport.
The SCF cycles are iterated until rms deviation of the potential is smaller than $10^{-7}$ Ry.
The calculated band structure of bulk graphite (with equilibrium structure) is shown in Fig.~5.
It is in general agreement with previous calculations that used norm-conserving,\cite{charlier94,holzwarth82} soft-transferable,\cite{schabel92} and ultrasoft\cite{furthfuller94} pseudopotentials, and also full-potential all-electron calculations with linearized augmented plane wave,\cite{jansen87} linear muffin-tin orbitals,\cite{ahuja95} and linear combination of Gaussian orbitals.\cite{boettger97}
The band energies are also in good agreement with experimental data (angle-resolved photoemission spectroscopy,\cite{arpes} infrared reflectance spectra,\cite{hanfland89,bellodi75} the angle-integrated photoemission spectroscopy,\cite{bianconi77} and angle resolved inverse photoemission spectroscopy\cite{aripes}), except the $\sigma_{2v}$ and $\sigma_{3v}$ bands at $\Gamma$-point.
As expected the LDA results underestimate the band gap. 
Differences among LDA calculations arise due to different $\bf k$-point sampling and energy cutoff used in the calculations.
However, the splittings at the $K$ point, which are relevant for the present study are among the best LDA results that reproduce the experimental data.

\begin{figure}
\vspace{-1.5cm}
\epsfxsize=3.25 truein
\centerline{\hspace{0cm}\epsfbox{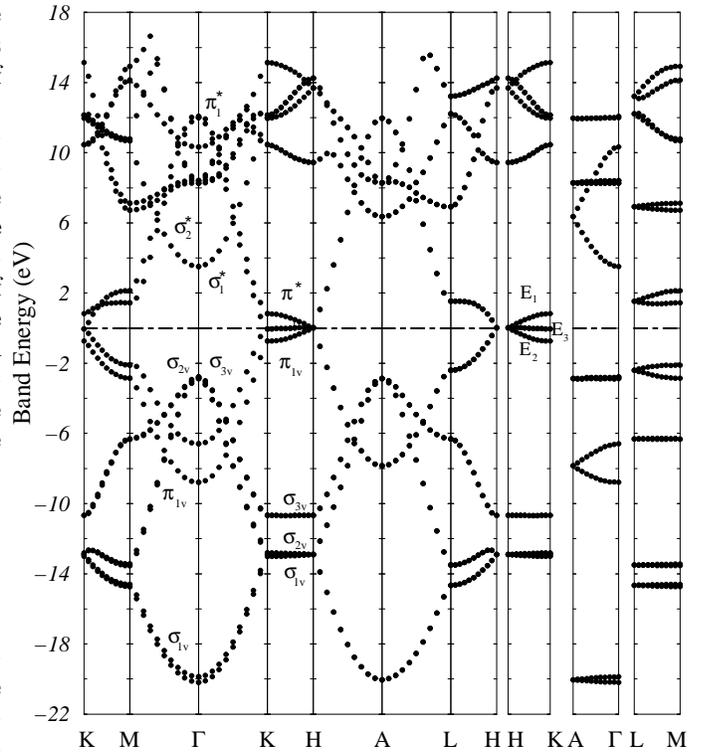}}
\vspace{.5cm}
\caption{The energy-band structure of graphite calculated by the SCF-pseudopotential method.
The equilibrium lattice parameters and  corresponding Brillouin zone are illustrated in Fig.~1.
The zero of the energy is taken at the Fermi level.}
\vspace{0cm}
\end{figure}

The energy band structure of graphene is essential for bands of bulk graphite.
Each graphene has four filled bands; three bands are due to the $\sigma$ bonds of the lateral $sp^2$ hybrid orbitals and one band is due to the dangling $p_z$ orbital forming delocalized bonds.
At the $\Gamma$ point these bands are labeled by $\sigma_{1v}$, $\pi_{1v}$, $\sigma_{2v}$, and $\sigma_{3v}$ in order of increasing energy. 
The empty bands are labeled by $\sigma_1^\ast$, $\sigma_2^\ast$, $\sigma_3^\ast$,  and $\pi_1^\ast$.
The band structure of graphite, that has four carbon atoms in the primitive unit cell (or two periodically repeating graphenes) can be viewed as if it consists of two superimposed graphene bands which are slightly shifted (split) owing to the weak interlayer coupling.
The dispersion of these bands are large ($5-10$ eV) when ${\bf k} $ is parallel to the (0001) plane owing to strong overlap of the orbitals in the relatively shorter lateral C-C distance.
However, for $\bf k$ perpendicular to  the (0001) plane (or ${\bf k} $ is parallel to the $z$ axis) the dispersion of the $\pi$ bands is small ($2-3$ eV); the $\sigma_{nv}$ bands are almost flat.
While the energy gap between the occupied and unoccupied bands is $\sim6$ eV at the $\Gamma$ point, it diminishes at the $K$ point and along the $KH$ direction where the bonding and antibonding bands join.
Accordingly, graphite is a semimetal.
Along the $KH$ direction $\pi$ bands originating from the bonding and antibonding combination of the $p_z$ orbitals located at $\alpha$ and $\alpha^\prime$ atoms (i.e $\pi_\alpha~{\rm and}~\pi_\alpha^\ast$) have significant dispersion  due to their coupling along the $z$ direction, whereas the dispersion of the $\pi_\beta$ and $\pi_\beta^\ast$ bands along the $KH$ direction is rather small.

The behavior of the $\pi$ bands along the $KH$ direction where they cross the Fermi level determine the Fermi surface and hence is essential for various physical properties of the bulk graphite.
The Fermi surface is generated by a $\bf k \cdot \bf p$ extension of these bands around the HKH axis (which is also known as the SWMc parametrization\cite{slonczewski58}), or by their Fourier expansion\cite{johnson73} of the LDA bands.
It includes six majority electron pockets located around the $H$ points, and 18 small pockets of minority electrons.
The majority carrier concentrations, $n_e\simeq 2.6\times 10^{18}~{\rm cm}^{-3}$ and $n_h\simeq 2.2\times 10^{18}~{\rm cm}^{-3}$ are reported from various Shubnikov-de Haas and de Haas-van Alphen experiments measuring the period of the magneto-resistance and magnetic susceptibility\cite{brandt88}.
The experimental values are reproduced by the LDA bands.

\begin{figure}
\vspace{-2cm}
\epsfxsize=3 truein
\centerline{\hspace{0.cm}\epsfbox{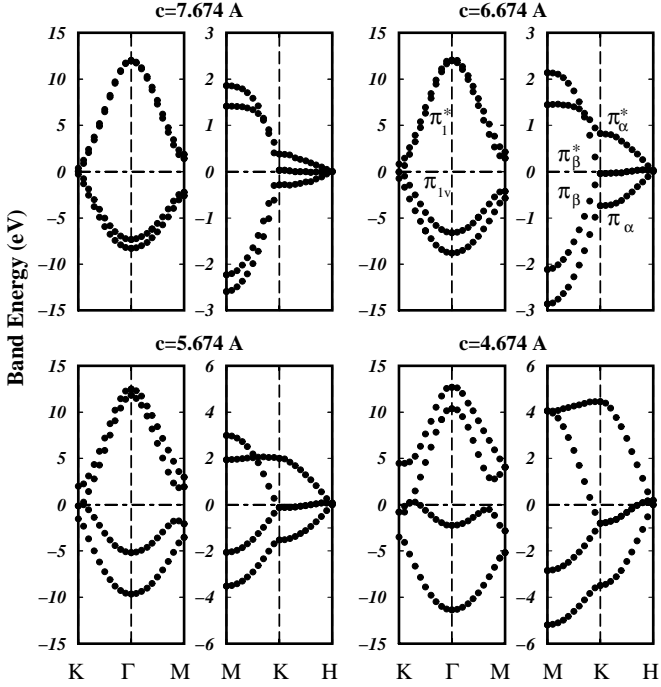}}
\vspace{.5cm}
\caption{The band structure of graphite is calculated for various values of the lattice parameter $c$ (or interlayer distance $d=c/2$) by using the SCF-pseudopotential method.
Only the $\pi$ bands along the relevant directions of the BZ are shown, and the zero of the energy is taken at the Fermi level.}
\vspace{0cm}
\end{figure}

\begin{figure}
\vspace{-.25cm}
\epsfxsize=3 truein
\centerline{\hspace{0cm}\epsfbox{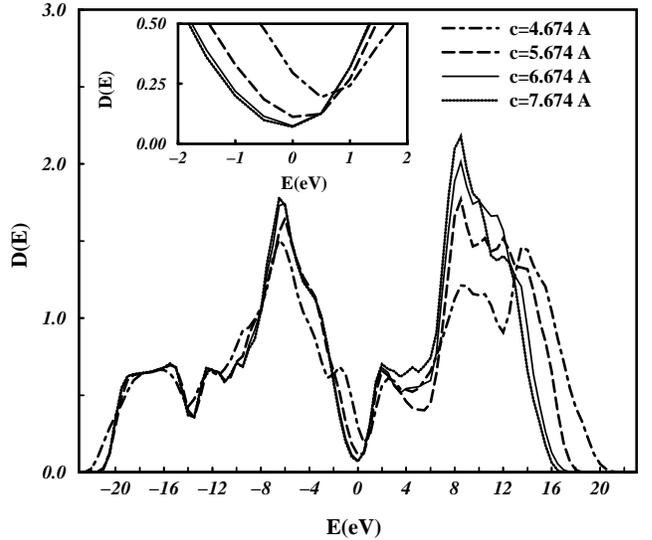}}
\vspace{-2.75cm}
\caption{The total density of states $D(E)$ is calculated for various values of the lattice parameter $c$.
The interlayer distance is $d=c/2$.
The density of states near $E_F$ is highlighted in the inset.}
\vspace{0cm}
\end{figure}

The effect of the uniaxial strain on the electronic structure can be explained by a simple tight-binding picture.
Considering only $p_z$ orbitals in the primitive unit cell located at $\alpha$, $\beta$, $\alpha^\prime$, and $\beta^\prime$ atoms (see Fig.~1), $\alpha$ and $\alpha^\prime$ atoms have significant overlap, whereas the overlap between $\beta$ and $\beta^\prime$ atoms is rather small.
Consequently, the $\pi_\alpha$ and $\pi_\alpha^\ast$ bands have to be dispersive, while the $\pi_\beta$ and $\pi_\beta^\ast$ bands stay almost flat along the $KH$ direction.
Furthermore, the dispersion of the $\pi_\alpha$ and $\pi_\alpha^\ast$ bands along the $KH$ direction increases with uniaxial compression since the overlap between $\pi_\alpha$ and $\pi_\alpha^\ast$ atoms increases.
The modification of the electronic structure under the uniaxial strain is illustrated by the {\it ab initio} bands shown in Fig.~6.
In comply with the above arguments, the $\pi_\alpha$ and $\pi_\alpha^\ast$ bands are strongly affected from the compressive strain.
The dispersion of the doubly degenerate $\pi_\beta$ bands in Fig.~6 is small but essential in determining the size of the electron pockets of the Fermi surface.
Note that already with compression the height of the BZ (i.e. $HKH$ distance in the $\bf k$ space) increases.
Furthermore, the overlap of the $\pi$ bands along $\Gamma K$ direction increases with increased compressive strain.
For $c\sim 4.67$ \AA~the $\pi_\beta$ band rises and touches the Fermi level along the $\Gamma M$ direction.

It becomes clear from the above discussion that the metallicity of the bulk graphite increases with increasing uniaxial compressive strain along the $z$ direction.
This situation can be revealed by examining the total density of states at the Fermi level $D(E_F)$ for various values of $c$.
Here we calculated the histogram of the state distribution by using $198~\bf k$ points in the irreducible wedge of the BZ and we then broaden it by a $0.5$ eV Gaussian convolution.
Our results are illustrated in Fig.~7.
It is seen that $D(E_F)$ may increase as much as $300\%$ upon $\sim30\%$ contraction of $c$.
Such an increase at $D(E_F)$ have important implications in the transport properties.\cite{mehrez98b}

\newpage
\begin{figure}
\vspace{-1.65cm}
\epsfxsize=5.25 truein
\centerline{\hspace{0cm}\epsfbox{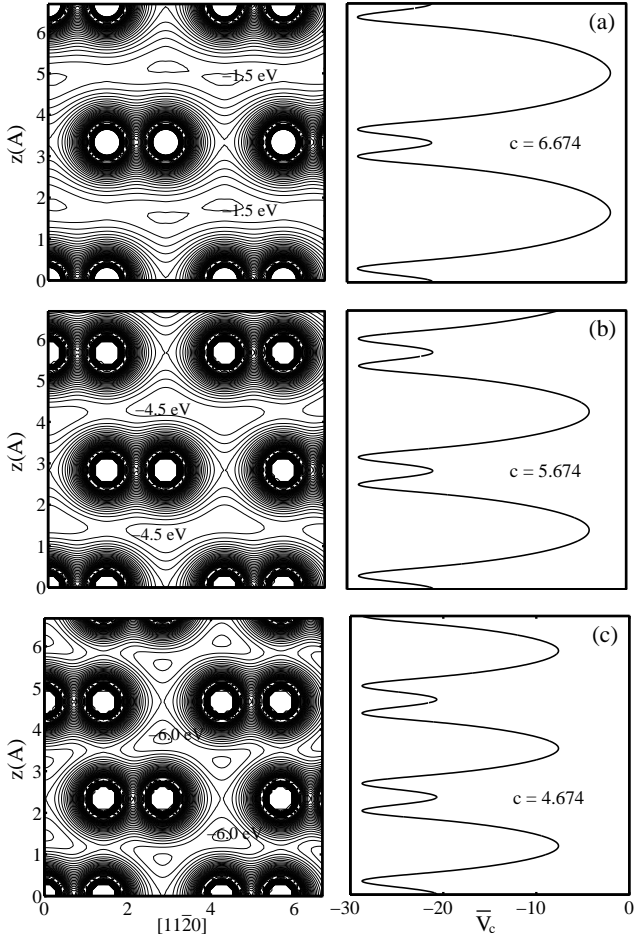}}
\vspace{-4.25cm}
\caption{Contour plots of the crystal potential on a perpendicular plane and the plane-averaged  crystal potential $\bar{V}_c(z)$ as a function of $c$. The contour interval is $1.5$ eV in each graph, and the indicated contour values are the maximum values with respect to the Fermi level. }
\vspace{-.25cm}
\end{figure}

Finally, we examine the variation of the crystal potential $V_c(\bf r)$ with a uniaxial compressive strain, i.e. with the interlayer spacing that is calculated selfconsistently.
The potential counterplots are shown in a relevant perpendicular plane in Fig.~8.
The planar averaged crystal potentials ($\bar{V}_c(z)=\frac{1}{\Omega}\int_\Omega V_c(x,y,z) dx~dy$, $\Omega$ being the cross section of the unit cell at the basal plane) are also shown in the same figure.
The crucial conclusion one can draw from these contour plots is that the potential between layers are lowered (becomes more attractive) with decreasing $c$.
Note that the crystal potential between layers is high and behaves as a barrier, but it is still below $E_F$.
In concluding this section, we emphasize that under the uniaxial compressive strain, that makes the interlayer distance smaller, the Fermi surface grows, $D(E_F)$ increases and $V_c(\bf r)$ between layers is lowered and becomes more attractive.
Thus the metallicity of graphite increases.
Because of the low carrier concentration and weak screening, such effects are expected to survive at the contact region.

\section{Conductance Through An Atomic-Size Contact}\label{sfour}

In the electron transport between the STM tip and sample one normally distinguishes two regimes, i.e. tunneling and ballistic.
The topological mode of STM (in which the separation is large enough to allow a vacuum barrier $\Phi$, and the bias voltage $V_B$ is small) operates in the tunneling regime.
Since the tunneling conductance is proportional to the local density of states of the sample at the center of the tip and at the Fermi energy, $\rho(E_F,\bf R_t)$,\cite{tersoff83} the tunneling current has been rather small in STM measurements on the graphite surface.
A significant value for the conductance can be achieved only at a small tip-sample separation, where the electronic contact is already established.
Earlier, it was found that, due to the tip-sample interaction, $\Phi$ collapses\cite{ciraci87} for $h\sim 3.5$ \AA~ and valence bands (normally below $E_F$) begin to overlap with the Fermi level in regions different from the $KH$ direction of the BZ.\cite{ciraci90}
Based on the {\it ab initio} electronic structure calculations it was also shown that the tip-induced states (TILS)(Ref.~6) form at the close proximity of the tip.
Actually, TILS are nothing but the states that are confined in the constriction and are precursors to the current transporting states.
In the present work, it is also shown that under the uniaxial stress (which is induced by the tip) the valence bands, which are normally below $E_F$, begin to cross the Fermi level.
The Fermi-level crossing may induce sudden changes in physical properties, in particular in the tunneling conduction.\cite{marti86}
Soler {\it et al.}\cite{soler86} were able to obtain only $G=0.05(2e^2/h)$ when the largest corrugation of line scans were achieved.
The significant tip-induced deformation suggests that the barrier collapses at the contact.
That the maximum conductance measured is much smaller than $2e^2/h$ disregards the opening of the first ballistic channel.
Such a situation, in which the ballistic conductance cannot set in despite the collapse of the barrier, i.e., $\Phi <E_F$, was pointed out earlier.\cite{ciraci90b}
The effective barrier $\Phi_{\rm eff}$, which may be greater than $E_F$ due to the size of the constriction, hinders the ballistic transport.\cite{ciraci90b}
On the other hand, it was argued that the flake of a dirt between the apex of the tip and the graphite surface increases the resistance and amplifies the corrugation\cite{pethica86}.
Nevertheless, apart from the spreading resistance, the conductance of a contact created by the sharp tip is generally smaller than $2e^2/h$.
This is due to the small density of states $D(E)$, available at $E_F$, since the current $I\propto \int_{E_F}^{E_F+eV_B} D_t(E)D(E+eV_B)T(E)dE$.
In this expression $D_t$ is the density of states of the tip, and $T(E)$ is the transmission coefficient.
The crystal potential between graphenes is also rather shallow, only $\sim2$ eV below $E_F$.
This may prevent a current transporting state (or a ballistic channel) from forming at the contact between two electrodes. 
As a result the transmission coefficient $T(E)$ becomes small.  
Another important feature that was not taken into account extensively in earlier works investigating the ballistic electron transport between two free-electron systems through a constriction is the actual topology of the Fermi surfaces.
While the Fermi surface of a metal tip can be taken (but not always) as a sphere, the Fermi surface of graphite is far from being similar to a sphere.
Since the narrow Fermi surface occurs at the corners of the hexagonal BZ, electrons at $E_F$ have a very large lateral momentum $\hbar{\bf k}_K$.
Even if the lateral momentum of the incoming electron cannot be conserved at the entrance of the contact, it should gain $\hbar {\bf k}_K$ after it passed to the graphite, which can be supplied by the acoustic phonons.
Therefore, the matching of the Fermi surfaces of the electrodes at both sides of the contact is important; the conductance measured at a low phonon population (or at low temperature) is expected to reveal interesting features.

\begin{figure}
\vspace{-.85cm}
\epsfxsize=4 truein
\centerline{\hspace{-.75cm}\epsfbox{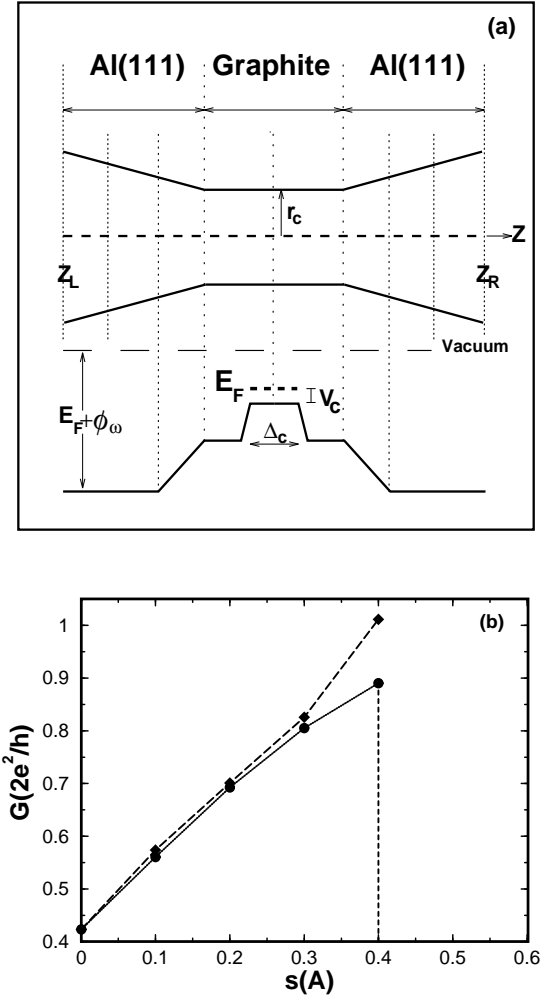}}
\vspace{-0cm}
\caption{(a)~A constriction model describing the point contact created by two Al(111) tip on both sides of a thin graphite slab.
The variation of the potential relative to the vacuum level is shown below.
The potential has a cylindrical symmetry.
(b)~Variation of the conductance with the displacement of the tip. 
The dotted and dashed curves correspond to cases (i) and (ii) explained in the text, respectively.
$\phi_\omega$ is the work function, and $r_c$, $\Delta_c$, $z_R$, and $z_L$ are defined in the text.}
\vspace{-.5cm}
\end{figure} 

Here we first present a qualitative discussion of the conductance through a contact created by a sharp tip.
As described in Sec.~\ref{stwo}, at a separation $h_s$ the surface layer of graphite is first attracted and jumps to the contact.
After this stage the separation between the tip and surface remain approximately unchanged for a significant displacement of the tip, unless a plastic deformation sets in (see Fig.~4).
Accordingly the current under constant $V_B$ does not increases significantly.
As $s$ increases, the tip presses the underlying graphite surface as shown in Fig.~3.
This decreases interlayer distance, lowers the interlayer potential  and increases the density of states at $E_F$ (see Figs.~6--8).
Even if $T(E)$ (that also increases with $s$) is assumed to be unchanged, the current and hence $G$ increase continuously with $s$ in the elastic region. 
However, as seen in Fig.~3, at certain range of $s$ (where $G$ reaches its highest value) the compressive stress exceeds a threshold value whereby the tip punctures the first layer by atomic rearrangements or by creating flakes at its close proximity.
This relaxes the strain in the layers.
As a result, the interlayer distance and the density of states are retrieved.
Since the tip faces now the second layer but with a larger separation, the conductance $G$ decreases suddenly.
As the tip first presses than punctures the graphite layers sequentially with increasing $s$, the conductance first increases from a low level to a high level and then falls again to a low level.
Due to the electron transport from the sides of the tip to the already punctured layers the low level value of $G$ may increase with increasing number of layers.
Nevertheless, $G(s)$ is expected to vary between low and high conductance values.
This situation may be different if a blunt (and flat) metal tip presses the graphite surface and forms a contact with large $A_c$.
Whether the quantum ballistic transport occurs for large contact area depends on the matching of the Fermi surfaces.

We now quantify the above discussion about $G(s)$ by calculating the conductance of the contact described in Fig.~9.
Two metal tips (Al oriented along the [111] direction) with a single atom at the apex make point contact from both surfaces of the graphite (0001) plane.
This contact can be described by a tapered constriction that has cylindrical symmetry and is connected to the free electron systems from both sides.
The Al electrodes are taken as free-electron systems with an electron density and $E_F$ appropriate for Al. 
The radius of the constriction through the graphite is determined from our earlier {\it ab initio} calculation dealing with Al tip and graphite surface\cite{ciraci90} and present calculations.
The crystal potential $V_c({\bf r})$ of graphite between two adjacent layers raises to $\sim 2$ eV below $E_F$.
This creates a saddle-point effect.\cite{ciraci87,buttiker90}
The potential of the constriction is described in Fig.~9(a).
After the mechanical contact is established, the displacement of the tip decreases the interlayer distance $c/2$ locally by applying a uniaxial compression.
This, in turn, induces changes in the electronic structure and the crystal potential.
For example, for the displacement $s\sim 0.4$ \AA~ (where the puncture sets in), the saddle point potential is lowered by $ 1.8$ eV for a three layer slab  shown in Fig.~9(a).
In this  simple constriction model, where the electrodes are represented by a free-electron gas system, we can consider only the variation of graphite potential and the radius of the constriction, i.e. $V_c(s)$ and $r_c(s)$.
Therefore, the calculation of conductance with $V_c(s)$ and $r_c(s)$ is expected to describe the essential features of the $G(s)$ curve.

To calculate conductivity, we used the recursion transfer matrix method proposed by Hirose 
and Tsukada;\cite{hirose95} and it was applied in similar types of calculations in Ref.~61.
In this approach, we replace our constriction shown in Fig.~9(a) by a periodically repeating supercell structure ($ 60\times60~{\rm a.u.}^2$) in the $xy$ plane defining the contact.
This allows us to express the {\sl j}{\it th} scattering state in terms of Laue expansion:
\begin{equation}
{\bf \Psi_j}({\bf r})={\rm e}^{i {\bf k_{||}\cdot {\bf r_{||}}}} \sum_l \psi_{lj}(z) {\rm e}^{i{{\bf G}_{||}^l\cdot {\bf r_{||}}}} . 
\end{equation}
Here $\bf k_{||}$ and $\bf G_{||}$ correspond to the wave vector in the BZ and the reciprocal lattice vector of the supercell in the $xy$ plane, respectively. 
The summation over $\bf G_{||}$ is truncated and a cutoff energy of 30 $eV$ was sufficient for convergence.
Deep in the Al structure, the wave function ${\bf \Psi}_j({\bf r})$ can be expressed as:
\begin{equation}
\left\{ \begin{array}{ll}
{\bf \Psi}_j({\bf r})&=\sum_{i}{\bf t}_{ij}
\left \{\begin{array}{lll}
{\rm e}^{i{\bf k}^{i}_{z}z}\\
{\rm e}^{-{\bf k}^{i}_{z}z}
\end{array}
\right \}{\rm e}^{i({\bf k}_{\|}+{\bf G}^{i}_{\|})\cdot {\bf r}_{\|}} \hspace*{.6cm}(z\ge z_R),\\
{\bf \Psi}_j({\bf r})&={\rm e}^{i{\bf k}^{j}_{z}\cdot z}{\rm e}^{i({\bf k}_{\|}+{\bf G}^{j}_{\|})\cdot {\bf r}_{\|}}\\
&+ \sum_{i}{\bf r}_{ij}
\left \{\begin{array}{lll}
{\rm e}^{-i{\bf k}^{i}_{z}\cdot z}\\
{\rm e}^{{\bf k}^{i}_{z}\cdot z}
\end{array}
\right \}{\rm e}^{i({\bf k}_{\|}+{\bf G}^{i}_{\|})\cdot {\bf r}_{\|}} \,\,\,\,\hspace*{.3cm}(z\le z_L),
\end{array}
\right.
\end{equation}
where $z_R$ and $z_L$ are the boundaries of the constriction. ${\bf t}_{ij}$ and ${\bf r}_{ij}$ are transmission and reflection coefficients, corresponding to states ${\bf k}^{i}_{z}$ and ${\bf k}^{j}_{z}$, and they are determined by using transfer-matrix method. 
We note that in this approach, the potential should be smooth and the constriction is divided into ``small strips" along the propagation direction (for recursive purposes), otherwise results diverge. 
To smooth out the potential at both ends of Al tips, we used a Fermi function of width $3$ a.u., and discretization width of $0.1$ a.u. was applied to obtain convergent results. 
In our model, we assume that $V_c(s)$ varies linearly (9.3 $\rightarrow$ 7.54 eV) as a function of $s$. Furthermore the mean width of the saddle-point potential $\Delta_c$ also decreases linearly with $s$. 
The radius of the constriction, $r_c(s)$, is also a difficult parameter to determine. 
Here we assume that $r_c(s)$ is uniform between the Al tips, and consider two limiting cases: (i) $r_c(s)=r_0=2.2$~\AA, i.e. the value corresponding to $h_s\sim2$~\AA~ at the electronic contact,\cite{ciraci89a} (ii) $r_c(s)=r_0 + 1.23s$, allowing a $50\%$ expansion of contact area before puncture occurs. 
Furthermore, linear response theory is assumed and we obtain:
\begin{equation}
G=\frac{2e^2}{h}\sum_{ij}|{\bf T}_{ij}|^2=\frac{2e^2}{h}\sum_{ij}(\delta_{ij}-|{\bf R}_{ij}|^2),
\end{equation}
where $i$ and $j$ run over all conducting states, ${\bf T}_{ij}={\bf t}_{ij}\times({\bf k}_{z}^{i}/{\bf k}_z^{j})^{1/2}$, and ${\bf R}_{ij}={\bf r}_{ij}\times({\bf k}_{z}^{i}/{\bf k}_z^{j})^{1/2}$.
We have considered only the case where ${\bf k}_{||}=0$, since the supercell area of the boundary condition is large and and hence the area of supercell BZ is negligible. 
We have applied both formulas in Eq.~(4) for determining $G$, because they serve as a good test of whether the results have converged. 
The calculated $G(s)$ curves are shown in Fig.~9(b).
As argued above, $G(s)$ increases with increasing $s$ until a plastic deformation (or puncturing of the graphene) sets in.
Once the graphene is punctured by the sharp tip, the conductance decreases suddenly.
For a sharp tip having a single atom at the apex, the radius of the constriction is normally small. 
In Fig.~9(b), $G$ attains a value smaller than $2e^2/h$, corresponding to case $(i)$ with $r_c=r_0$. 
For case (ii), where $r_c(s)$ is allowed to expand, a ballistic channel is opened before the puncturing.
We also note that $\Delta G / \Delta s$ of actual graphite is expected to be much smaller than that of Fig.~8(c), since the same level of compressive strain in Fig.~8(c) requires much larger $s$ in the multilayer graphite.
For the constriction described in Fig.~9(a), $s\simeq \Delta_c$.

\section{Conclusions}

In this work, based on the atomistic simulation of the contact and {\it ab initio} electronic structure calculations of graphite under uniaxial compressive strain  we showed the following: 
(i) The graphenes are punctured sequentially by a sharp tip pressing the surface. 
(ii) The tip induced elastic deformation between two consecutive puncture increases the density of states at the close proximity of the tip, and lowers the crystal potential: 
Hence, the metallicity of graphite increases with uniaxial compressive strain of basal planes.
(iii) Accordingly the conductance between two consecutive puncture increases, but falls suddenly upon the onset of a new puncture.
This is a behavior different from the usual metal contacts.

\section{acknowledgment}
We thank K. Nordlund for providing his empirical potential data.

\end{document}